\documentclass[12pt,preprint]{aastex}

\slugcomment{}

\shortauthors{Stanek et al.}
\shorttitle{Spectroscopic Supernova Signature in GRB\,030329}

\begin{document}


\title{Spectroscopic Discovery of the Supernova 2003dh  Associated
with GRB\,030329\altaffilmark{1}}

\author{K.~Z.~Stanek\altaffilmark{2},
T.~Matheson\altaffilmark{2},
P.~M.~Garnavich\altaffilmark{3},
P.~Martini\altaffilmark{4},
P.~Berlind\altaffilmark{5},
N.~Caldwell\altaffilmark{2},
P.~Challis\altaffilmark{2},
W.~R.~Brown\altaffilmark{2},
R.~Schild\altaffilmark{2},
K.~Krisciunas\altaffilmark{6,7},
M.~L.~Calkins\altaffilmark{5},
J.~C.~Lee\altaffilmark{8},
N.~Hathi\altaffilmark{9},
R.~A.~Jansen\altaffilmark{9},
R.~Windhorst\altaffilmark{9},
L.~Echevarria\altaffilmark{9},
D.~J.~Eisenstein\altaffilmark{8},
B.~Pindor\altaffilmark{10},
E.~W.~Olszewski\altaffilmark{8},
P.~Harding\altaffilmark{11},
S.~T.~Holland\altaffilmark{3},
D.~Bersier\altaffilmark{2}}

\altaffiltext{1}{Based on data from the MMTO 6.5m telescope, the
Magellan 6.5m Clay telescope, and the FLWO 1.5m telescope}

\altaffiltext{2}{Harvard-Smithsonian Center for Astrophysics, 60
Garden Street, Cambridge, MA 02138; kstanek, tmatheson, ncaldwell,
pchallis, wbrown, rschild, dbersier@cfa.harvard.edu}

\altaffiltext{3}{Department of Physics, University of Notre Dame, 225
Nieuwland Science Hall, Notre Dame, IN 46556;
pgarnavi@miranda.phys.nd.edu, sholland@nd.edu}

\altaffiltext{4}{Carnegie Observatories, 813 Santa Barbara Street,
Pasadena, CA 91101; martini@ociw.edu}

\altaffiltext{5}{Smithsonian Institution, F.~L.~Whipple Observatory,
670 Mt.~Hopkins Road, P.O.~Box 97, Amado, AZ 85645; pberlind,
mcalkins@cfa.harvard.edu}

\altaffiltext{6}{Las Campanas Observatory, Casilla 601, La Serena, Chile}

\altaffiltext{7}{Cerro Tololo Inter-American Observatory, Casilla 603,
La Serena, Chile; kevin@ctiosz.ctio.noao.edu}

\altaffiltext{8}{Steward Observatory, University of Arizona, 933
N.~Cherry Ave., Tucson, AZ 85718; janice@pompelmo.as.arizona.edu,
eisenste@cmb.as.arizona.edu, edo@adansonia.as.arizona.edu}

\altaffiltext{9}{Department of Physics and Astronomy, Arizona State
University, Tempe, AZ 85287-1504; Nimish, Rolf.Jansen,
Rogier.Windhorst, Luis.Echevarria@asu.edu}

\altaffiltext{10}{Princeton University Observatory, Princeton, NJ 08544;
pindor@astro.princeton.edu}

\altaffiltext{11}{Department of Astronomy, Case Western Reserve
University, 10900 Euclid Avenue, Cleveland, OH 44106;
harding@billabong.astr.cwru.edu}

\begin{abstract}

We present early observations of the afterglow of the Gamma-Ray Burst
(GRB) 030329 and the spectroscopic discovery of its associated
supernova SN\,2003dh.  We obtained spectra of the afterglow of GRB
030329 each night from March 30.12 (0.6 days after the burst) to
April~8.13 (UT) (9.6 days after the burst). The spectra cover a
wavelength range of 350 nm to 850 nm. The early spectra consist of a
power-law continuum ($F_{\nu}\propto \nu^{-0.9}$) with narrow emission
lines originating from HII regions in the host galaxy, indicating a
low redshift of $z=0.1687$.  However, our spectra taken after 2003
Apr.~5 show broad peaks in flux characteristic of a supernova.
Correcting for the afterglow emission, we find the spectrum of the
supernova is remarkably similar to the type~Ic `hypernova' SN\,1998bw.
While the presence of supernovae have been inferred from the light
curves and colors of GRB afterglows in the past, this is the first
direct, spectroscopic confirmation that a subset of classical
gamma-ray bursts originate from supernovae.

\end{abstract}

\keywords{galaxies: distances and redshifts --- gamma-rays: bursts ---
supernovae: general --- supernovae: individual (SN\,2003dh)}

\section{Introduction}

The origin of gamma-ray bursts (GRB) has been a mystery since their
discovery in the 1960's.  It has only been since the BeppoSAX
satellite (Boella et al.~1997) began providing rapid, accurate
localization of several bursts per year that it has it been possible
to study these events and their afterglows in detail. Optical
observations of afterglows (e.g.~GRB\,970228: Groot et al.~1997; van
Paradijs et al.~1997) have allowed redshifts to be measured for a
number of GRBs (e.g.~GRB 970508: Metzger et al.~1997), providing
definitive proof of their cosmological origin.  GRB\,980425 was likely
associated with supernova 1998bw and this was the first direct
evidence that at least some GRB result from the core collapse of
massive stars (Galama et al.~1998).  However the isotropic energy of
that burst was 10$^{-3}$ to 10$^{-4}$ times weaker (Woosley, Eastman,
\& Schmidt 1999) than classical cosmological GRB which placed it in a
unique class.  Indirect evidence of the connection between GRB and
massive stars has come from studies of the location of GRB in their
host galaxies (e.g.~Holland \& Hjorth 1999) and statistics on the
types of galaxies that host GRB (e.g.~Hogg \& Fruchter 1999).
Chevalier \& Li (2000) have shown that the afterglow properties of
some GRB are consistent with a shock moving into a stellar wind formed
from a massive star.

Direct evidence of a classical GRB/supernova connection has been
difficult to obtain because the typical redshift of a GRB is $z\sim
1$, meaning even powerful supernovae would peak at $R > 23$ mag.  A
number of GRB have shown late deviations from a power-law decline
(e.g.~GRB\,980326, Bloom et al.~1999) which are suggestive of a
supernova peaking a few weeks after the burst. At $z=0.36$,
GRB\,011121/SN2001ke was a relatively nearby burst which showed a
late-time bump and color changes consistent with a supernova
(Garnavich et al.~2003a; Bloom et al.~2002). That burst was indeed the
best evidence to date that classical, long gamma-ray bursts are
generated by core-collapse supernovae, but it lacked a clear
spectroscopic detection of a supernova signature.  Detection of such a
signature is reported in this paper for the GRB 030329.

The extremely bright GRB\,030329 was detected by the FREGATE, WXM, SXC
instruments aboard HETE II at 11:14:14.67 UT on 2003 March~29
(Vanderspek et al.~2003). The burst falls in the `long' category with
a duration of $>25$ seconds.  Peterson \& Price~(2003) and
Torii~(2003) reported discovery of a very bright ($R\sim13$), slowly
fading (Uemura 2003) optical transient (OT), located at
$\alpha$~=~$10^{\rm h}44^{\rm m}50\fs0$,
$\delta$~=~$+21\arcdeg31\arcmin17\farcs8$ (J2000.0), and identified
this as the GRB optical afterglow. Due to the brightness of the
afterglow and its slow decay, photometric observations were extensive,
making it one of the best-observed afterglows.

The afterglow was also very bright in X-rays (Marshall \& Swank 2003),
radio (Berger, Soderberg \& Frail 2003), sub-millimeter (Hoge et
al.~2003) and infrared (Lamb et al.~2003).  Using archival data Blake
\& Bloom (2003) have put a $3\sigma$ upper limit of R=22.5 on the
brightness of the host.  Martini et al.~(2003) were the first to
report optical spectroscopy of the afterglow. Because the very bright
OT overwhelmed the emission from the faint host galaxy, only a single
emission line was detected and confirmed by Della Ceca et al.~(2003).
Suggesting that this line may be due to [OII] provided a redshift of
$z\approx 0.5$.  However, a high-resolution VLT spectrum by Greiner et
al.~(2003) revealed additional emission and also absorption lines
which fixed the redshift at a very low $z=0.1685$. This was later
confirmed by Caldwell et al.~(2003), making GRB\,030329 the second
nearest burst overall (GRB\,980425 is the nearest at z=0.0085) and the
classical burst with the lowest known redshift.

\section{Observations}

From the moment the low redshift for the GRB\,030329 was announced
(Greiner et al.~2003), we started organizing a campaign of
spectroscopic and photometric follow-up of the afterglow and later the
possible associated supernova. Spectra of the OT associated with GRB
030329 were obtained over many nights with the 6.5m MMT telescope, the
1.5m Tillinghast telescope at the F.~L.~Whipple Observatory (FLWO),
and the Magellan 6.5m Clay telescope.  The spectrographs used were the
Blue Channel (Schmidt et al.~1989) at the MMT, FAST (Fabricant et
al.~1998) at the FLWO 60'' telescope, and LDSS2 (Mulchaey 2001) at
Magellan.  The observations were reduced in the standard manner with
IRAF\footnote{IRAF is distributed by the National Optical Astronomy
Observatory, which is operated by the Association of Universities for
Research in Astronomy, Inc., under cooperative agreement with the
National Science Foundation.} and our own routines.  Spectra were
optimally extracted (Horne 1986).  Wavelength calibration was
accomplished with HeNeAr lamps taken immediately after each OT
exposure.  Small-scale adjustments derived from night-sky lines in the
OT frames were also applied.  We used Feige 34 (Stone 1977) and HD
84937 (Oke \& Gunn 1983) as spectrophotometric standards.  We
attempted to remove telluric lines using the well-exposed continua of
the spectrophotometric standards (Wade \& Horne 1988; Matheson et
al.~2000).  The spectra were in general taken at or near the
parallactic angle (Filippenko 1982).  The relative fluxes are thus
accurate to $\sim$ 5\% over the entire wavelength range.  Figure
\ref{grb-all} shows a subset of our spectra.  The initial report on
some of these data was presented by Martini et al.~(2003) and Caldwell
et al.~(2003).  Circulars by Matheson et al.~(2003a), Garnavich et
al.~(2003b), Matheson et al.~(2003b) and Garnavich et al.~(2003c)
reported the first detection of supernova associated with GRB\,030329
(see next section).

\section{Results}

\begin{figure}
\epsscale{1.0}
\plotone{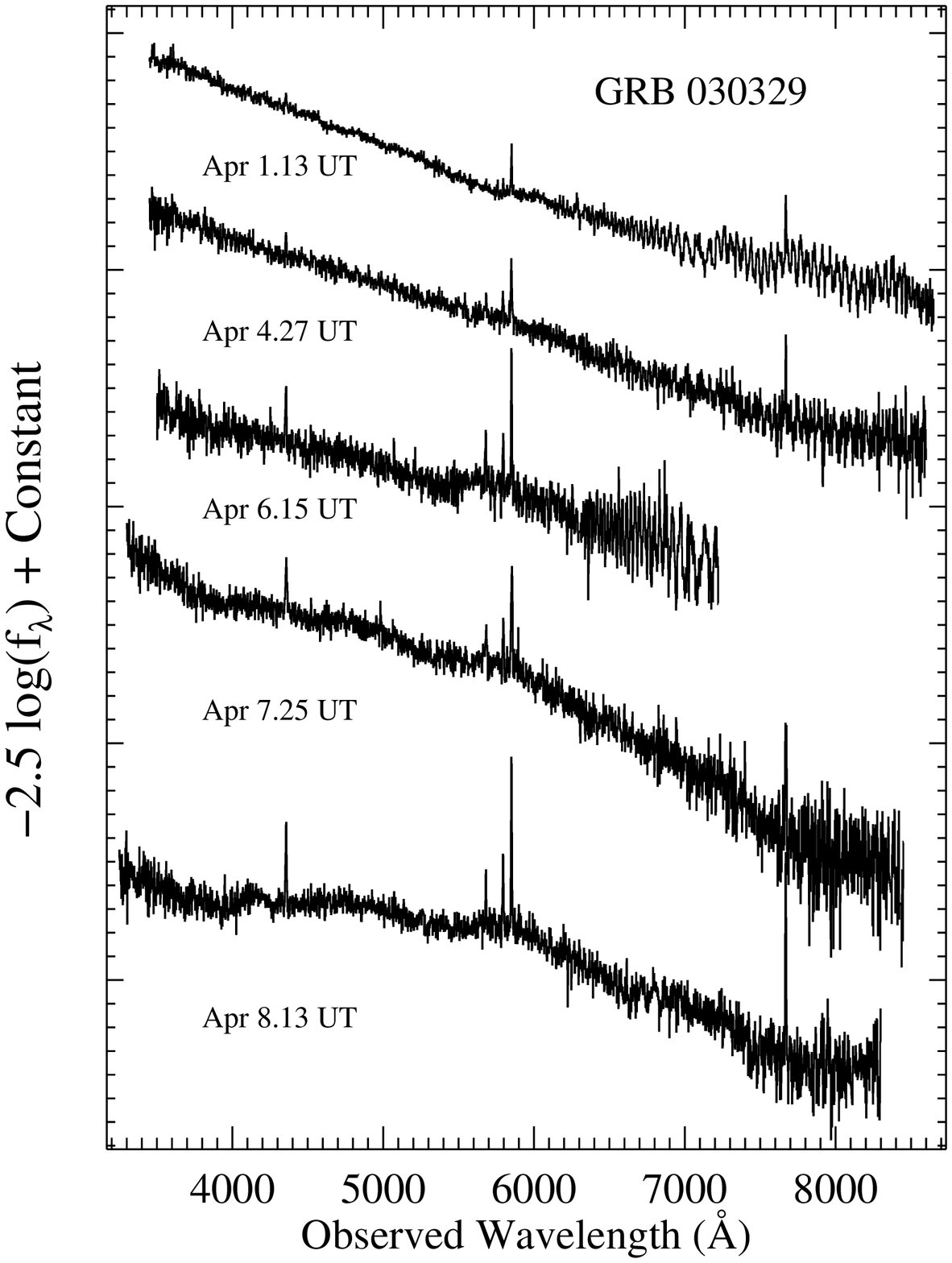}
\caption{Evolution of the GRB\,03029/SN\,2003 spectrum, from April~1.13 UT
(2.64 days after the burst), to April 8.13 UT  (9.64 days after
the burst). The early spectra consist of a power-law continuum
($F_{\nu}\propto \nu^{-0.9}$) with narrow emission lines originating
from HII regions in the host galaxy at a redshift of z=0.168. Spectra
taken after Apr.~5 show the development of broad peaks in the spectra
characteristic of a supernova.
\label{grb-all}}
\end{figure}

The brightness of the OT and its slow and uneven rate of decline, with
episodes of increased brightness (well documented by many GCN
circulars) allowed us to observe the OT every night since the GRB
event (so far until Apr.~9 UT), thus providing a unique opportunity to
look for spectroscopic evolution over many nights.  The early spectra
of the OT of GRB\,030329 (top of Figure \ref{grb-all}) consist of a
power-law continuum typical of GRB afterglows, with narrow emission
features identifiable as H$\alpha$, [OIII], H$\beta$ and [OII] at $z =
0.1687$ (Greiner et al.~2003; Caldwell et al.~2003) probably from HII
regions in the host galaxy.  These lines can be used to estimate the
star formation rate within the spectrograph slit (Kennicutt 1998).
Preliminary analysis of the [OII] flux suggests a low star formation
rate of $\sim 0.1\;M_{\odot}\;yr^{-1}$ (Caldwell et al.~2003), but a
more detailed analysis is possible after the afterglow has faded.

A fit to the early spectra provides a power-law index of $\beta=-0.94
+/- 0.01$ for April~1 and $-0.93 +/- 0.01$ for April~4 (statistical
errors only). This is consistent with the spectral slope of $-0.94$
found using SDSS photometry taken March~31/April~1 (Lee et al.~2003).
Correcting for low Galactic extinction of $E(B-V)=0.025$ (Schlegel et
al.~1998) lowers the slope to $-0.85$ which is a typical spectral
index for GRB afterglows (e.g.~Stanek et al.~2001).

Beginning April~6, our spectra showed the development of broad peaks
in flux, characteristic of a supernova. The broad bumps are seen at
approximately 5000\AA\ and 4200\AA\ (rest frame). At that time, the
spectrum of GRB\,030329 looked similar to that of the peculiar type Ic
SN\,1998bw a week before maximum light (Patat et al.~2001). Over the
next few days the SN features became more prominent as the afterglow
faded and the SN brightened towards maximum.

\begin{figure}
\epsscale{0.8}
\plotone{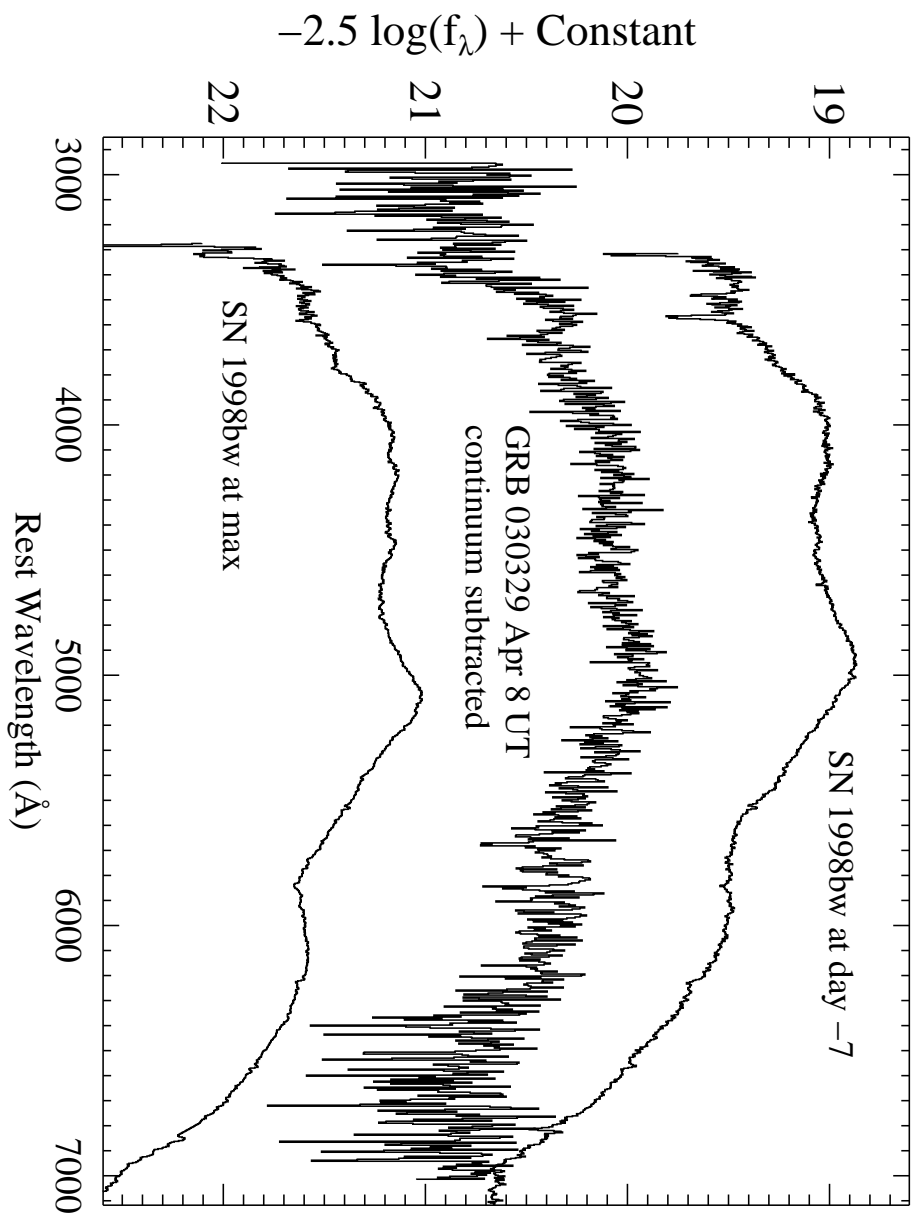}
\caption{MMT spectrum of April~8 with the smoothed MMT spectrum of
April~1 scaled and subtracted.  The residual spectrum shows broad
bumps at approximately 5000\AA\ and 4200\AA\ (rest frame), which is
similar to the spectrum of the peculiar type Ic SN\,1998bw a week
before maximum light (Patat et al.~2001). The match is not as good for
SN\,1998bw at maximum light, especially at the red end of the spectrum.
\label{grb-diff}}
\end{figure}

To discern the spectrum of the SN component, we assume that the
spectral slope of the afterglow light did not evolve significantly in
time.  This allows us to subtract the afterglow dominated spectrum
obtained on April~4 UT (see Figure \ref{grb-all}) from the supernova
dominated spectrum of April~8. The resulting spectrum is shown in
Fig~\ref{grb-diff}. For comparison, spectra of SN\,1998bw at maximum
and a week before maximum are also shown (Patat et al.~2001). The
similarities of the GRB\,030329 supernova to SN\,1998bw are striking
while the match to other `hypernovae' such as SN\,1997ef (Iwamoto et
al.~2000) and SN\,2002ap (Mazzali et al.~2002) is not as good. The
primary difference is that the feature around 4400~\AA\ (rest frame)
in GRB\,030329 supernova and SN\,1998bw is very broad while in
SN\,1997ef and SN\,2002ap the feature is sharp and well defined.  This
is likely an indication that the expansion velocities in 1998bw and
GRB\,030329 supernova are significantly higher than in the other two
events.

The evolution of the GRB\,030329/SN\,2003dh is still ongoing.  In a
future paper (Matheson et al.~2003, in preparation) we will discuss in
more detail the properties of the afterglow of GRB\,030329, in itself
a very unusual event, and long-term spectroscopic and photometric
evolution of the SN\,2003dh.  With this future paper we will also
release our data for this GRB/SN via {\tt anonymous ftp}.

We have shown convincing spectroscopic evidence that a supernova was
lurking beneath the optical afterglow of the classical, long gamma-ray
burst 030329. The supernova spectrum was very similar to the type Ic
`hypernova' SN\,1998bw which was associated with the intrinsically
weak GRB\,980425.  With two confirmed cases, it is tempting to link
only type~Ib/c events with GRB. However, the type~IIn hypernova 1997cy
(Germany et al. 2000) may have triggered a GRB and the blue color of
SN~2001ke near maximum was consistent with some hydrogen rich events
(Garnavich et al. 2003a).

The 4400\AA\ feature in the spectrum of GRB\,030329 is significantly
broader than seen in `hypernovae' SN\,1997ef and SN\,2002ap, neither
of which was clearly associated with a GRB event.  While the presence
of supernovae have been inferred from the light curves and colors of
GRB afterglows in the past, this is the first direct, spectroscopic
confirmation that some and maybe all classical gamma-ray bursts
originate from supernovae.

\acknowledgments

We would like to thank the staffs of the MMT, Las Campanas and the
FLWO observatories. Thanks to A.~Milone for helping with the MMT
spectra.  PMG acknowledges the support of NASA/LTSA grant NAG5-9364.
PM was supported by a Carnegie Starr Fellowship.  JCL acknowledges
financial support from NSF grant AST-9900789.  DJE was supported by
NSF grant AST-0098577 and by an Alfred P. Sloan Research Fellowship.
BP is supported by NASA grant NAG5-9274. EWO was partially supported
by NSF grant AST 0098518.

\end{document}